\documentclass{desyproc}

\begin{document}
\title{QCD Axion and Dark Energy}

\author{{\slshape Jihn E. Kim } \\
Department of Physics, Kyung Hee University, Seoul 130-701, Korea, and\\
Department of Physics, Seoul National University, Seoul 151-747, Korea}

\contribID{Kim\_Jihn}

\desyproc{DESY-PROC-2014-XX}
\acronym{Patras 2014} 

\newcommand{\dis}[1]{\begin{equation}\begin{split}#1\end{split}\end{equation}}

\newcommand{\tev}{\,\textrm{TeV}}
\newcommand{\gev}{\,\textrm{GeV}}
\newcommand{\meV}{\,\mathrm{MeV}}
\newcommand{\keV}{\,\mathrm{keV}}
\newcommand{\eV}{\,\mathrm{eV}}
\newcommand{\Mp}{M_{\rm P}}
\newcommand{\Mpt}{$M_{\rm P}$}
\newcommand{\Mgt}{$M_{\rm GUT}$}
\newcommand{\Mg}{M_{\rm GUT}}
\newcommand{\Uone}{U(1)$_{\rm gl}$}
\newcommand{\Ude}{U(1)$_{\rm de}$}
\newcommand{\Uga}{U(1)$_{\rm ga}$}
\newcommand{\UPQ}{U(1)$_{\rm PQ}$}
\newcommand{\tut}{$t_{\rm U}$}
\newcommand{\tu}{t_{\rm U}}
\newcommand{\ie}{{\it i.e.~}}
\newcommand{\etal}{{\it et al.}\,}

\doi  

\maketitle

\begin{abstract}
String allowed discrete symmetry is the mother of acceptable effective global symmetries at low energy. With this philosophy, we discuss dark energy, QCD axion, and inflation, and speculate some implications of the recent BICEP2 data.
\end{abstract}

\section{Pseudo-Goldstone Bosons from Discrete Symmetries}

The ongoing QCD axion detection is based on the bosonic coherent motion (BCM). The QCD axion can be a fundamental pseudoscalar or a composite one \cite{KimAxComp}. But, after the discovery of the fundamental Brout-Englert-Higgs (BEH) boson, a fundamental QCD axion gain much more weight. This leads to a BEH portal to the high energy scale to the axion scale or even to the standard model (SM) singlets at the grand unification (GUT) scale. Can these singlets explain both dark energy (DE) and cold dark matter (CDM) in the Universe? Because the axion decay constant $f_a$, signifying the axion window, can be in the intermediate scale, axions
can live up to now ($m_a <24\, \eV$) and constitute DM of the
Universe. In this year of a GUT scale VEV,  can these explain  the inflation finish in addition to  DE and CDM?

\begin{figure}[!b]
\centerline{\includegraphics[width=0.35\textwidth]{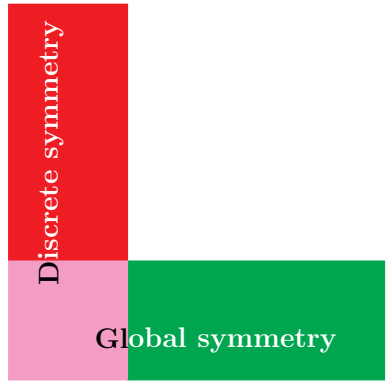}
 }
\caption{Terms respecting discrete and global symmetries.}\label{Fig:discrete}
\end{figure}

 But we have to worry about the quantum gravity effects which are known to break global symmetries, especially via the Planck scale  wormholes.   To resolve this dilemma, we can think of two possibilities of discrete symmetries below \Mpt\,\cite{Kim13worm}: (i) The discrete symmetry arises as  a part  of  a gauge symmetry,
and (ii) The string selection rules directly give the discrete symmetry. So, we will consider discrete gauge symmetries allowed in string compactification. Even though the Goldstone boson
directions in spontaneously broken gauge symmetries are flat, the Goldstone boson directions of spontaneously broken {\em global} symmetries are not flat, \ie global symmetries are always {\em approximate}. The question is what is the degree of the {\em approximateness}. In Fig. \ref{Fig:discrete}, we present a cartoon separating effective terms according to string-allowed discrete
symmetries. The terms in the  vertical column represent exact symmmetries such as gauge symmetries and string allowed discrete symmetries. If we consider a few terms in the lavender part, we can consider a {\em global symmetry}. The global symmetry is broken by terms in the red part.

The most studied global symmetry is the Peccei-Quinn (PQ) symmetry \UPQ. For \UPQ, the dominant breaking term is the QCD anomaly term $(\overline{\theta}/32\pi^2)G_{\mu\nu}\tilde{G}^{\mu\nu}$ where $G_{\mu\nu}$ is the gluon field strength. Since  $\overline{\theta}$ gives a neutron EDM (nEDM) of order $10^{-16}\overline{\theta}\,e$cm, the  upper bound on nEDM restricts $|\overline{\theta}|<10^{-10}$. ``Why is $\overline{\theta}$ so small?" is the strong CP problem \cite{KimRMP}.

Axion osmology  was  started in 1982--1983. One key point I want to stress here is the axion domain wall (DW) problem \cite{Vilenkin82,Sikivie82}. But the recent numerical study \cite{Kawasaki12} of axion creation by decay of axionic DWs is too large, which gave a strong lower bound on the axion mass \cite{Marsh14} from the BICEP2 data \cite{Larger}. For the DW problem from string, there is a solution of Choi-Kim \cite{ChoiKimDW85}, which has been recently used for the anomalous U(1) from string \cite{KimDW14,BarrKim14} in view of the GUT scale inflation implied by the BICEP2 data.  A smoking gun for the axion CDM is detecting an oscillating $\overline{\theta}$. In this spirit, an
oscillating nEDM was suggested to be measured 20 years ago \cite{Hong90,Hong91}, and it got renewed  interests recently \cite{Hong14,Asimina14}. A guideline for the axion-photon-photon coupling is a string calculation, $c_{a\gamma\gamma}\simeq 0.91$ \cite{Kimagg14} in the ${\bf Z}_{12-I}$ model \cite{HuhKK09}.

\begin{figure}[!b]
\centerline{\includegraphics[width=0.5\textwidth]{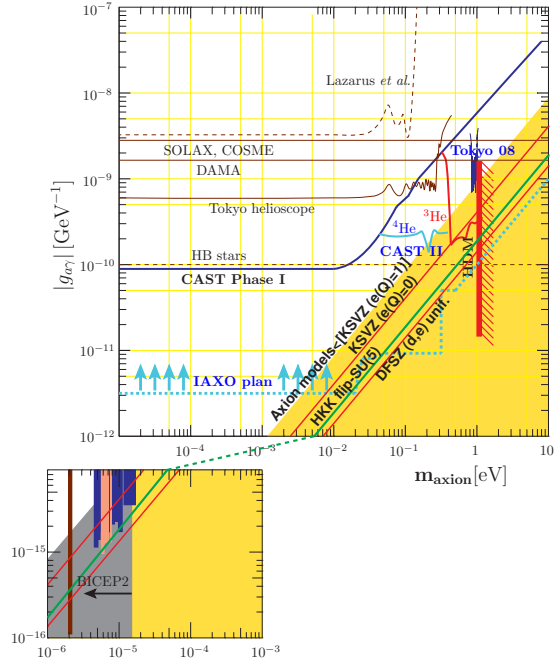}
 }
\caption{The $g_{a\gamma}$ coupling vs. axion mass. The green line is a string calculation \cite{Kimagg14}.}\label{Fig:data}
\end{figure}

\section{Dark Energy from \Ude}

The axion oscillation is just one example of  BCM, classified as {\bf BCM1} in \cite{KimYannShinji}.
The QCD axion naturally arises if one tries to interpret the DE scale by spontaneously breaking an {\em approximate} U(1) global symmetry, \Ude. A recent calculation of the cosmic axion density gives the axion window, $10^{9\,}\gev<f_a<10^{12\,}\gev$, including the anharmonic term carefully with the new data on light quark masses \cite{Bae08}. It is known that string axions from $B_{MN}$ have GUT scale decay constants \cite{ChoiKimst85}. Therefore, the QCD axion from string theory is better to arise from matter fields originating from 10-dimensional adjoint representation of E$_8\times$E$_8'$ \cite{KimPRL14}.
Note that the global symmetry violating terms belong to the red part in Fig. \ref{Fig:discrete}. For the QCD axion, the dominant breaking is by the QCD anomaly term. The QCD anomaly term is too large to account for the DE scale of $10^{-46\,}\gev^4$.  Because the BEH scalar is known to be a fundamental one, we extend the fundamentality to the SM singlet scalars also. These singlets must couple to $H_u H_d$, \ie a kind of BEH portal. So, to relate the DE scale to a pseudo-Goldstone boson from \Ude, we must forbid the coupling to the QCD anomaly term, which can be done only by introducing another {\em global} U(1), \ie to remove the \Ude-$G$-$G$
anomaly with $G=$ QCD. Namely, another U(1) must be introduced so that one linear combination is free
of the  QCD anomaly. Thus, introduction of two global symmetries is inevitable to interpret the DE scale in terms of a potential of a pseudo-Goldstone boson \cite{KimNilles14}. Let us call these two global symmetries as \UPQ~and \Ude.

The breaking scale of \Ude~is trans-Planckian \cite{Carroll98}. With \UPQ\,and \Ude, one can construct a DE model from string compactification \cite{KimJKPS14}. Using the SUSY language,
the discrete and global symmetries below $\Mp$ are the consequence of the full superpotential $W$. So, the exact symmetries related to string compactification are respected by the full $W$, \ie the vertical column of Fig. \ref{Fig:discrete}. Considering only the $d=3$ superpotential $W_3$,  we can consider an
approximate PQ symmetry.
For the MSSM interactions supplied by R-parity, one needs to know all the SM singlet spectrum. We need ${\bf Z}_2$ for a WIMP candidate.

The DE potential height is so small, $10^{-46\,}\gev^4$, that the needed discrete symmetry breaking term of Fig. \ref{Fig:discrete} must be small, implying the discrete symmetry is of high order. Now, we have a scheme to explain both 68\% of DE and 27\% of CDM via approximate {\em global} symmetries. With SUSY, axino may contribute to CDM also \cite{Baer14}.

A typical example for the  discrete symmetry is ${\bf Z}_{10\,R}$ as shown in \cite{KimJKPS14}. The
${\bf Z}_{10\,R}$  charges descend from a gauge U(1) charges of the string compactification \cite{HuhKK09}. The U(1) defining terms, \ie the lavender part of Fig \ref{Fig:discrete}, is  $W_6$. Then, the height of the potential is highly suppressed and we can obtain $10^{-47\,}\gev^4$, without the gravity spoil of the global symmetry. In this scheme with BEH portal, we introduced three VEV scales, TeV scale for $H_uH_d$, the GUT scale \Mgt~for singlet VEVs, and the intermediate scale for the QCD axion. The other fundamental scale is $\Mp$. The trans-Planckian decay constant $f_{\rm de}$ can be a derived scale \cite{KNP05}.

Spontaneous breaking of \Ude~is via a Mexican hat potential with the height of $\Mg^4$.
A byproduct of this Mexican hat potential is the hilltop inflation with the height of $O(\Mg^4)$.
It is a small field inflation, consistent with the recent Planck data.

\section{Gravity Waves  from U(1)$_{\rm de}$ and its Implications}

With the surprising report from the BICEP2 group on a large tensor-to-scalar ratio $r$  \cite{Larger}, we must consider the above hilltop inflation whether it leads to appropriate numbers on $n_s, r$ and the e-fold number $e$ or not. With two U(1)'s, the large trans-Planckian $f_{\rm de}$ is not spoiled by the intermediate PQ scale $f_a$ because the PQ scale just adds to the $f_{\rm de}$ decay constant only by a tiny amount.

Inflaton potentials with almost flat one near the origin, such as the Coleman-Weinberg type new inflation, were the early attempts for inflation. But any models can lead to inflation if the potential is flat enough as in the {\em chaotic inflation} \cite{Linde83}. A single field chaotic inflation survived now is the $m^2\phi^2$ scenario {\em chaotic inflation}. To shrink the field energy much smaller than $\Mp^4$, a {\em natural inflation} (mimicking the axion-type minus-cosine potential) has been introduced \cite{Freese90}. If a large $r$ is observed, Lyth noted that the field value $\langle\phi\rangle$ must be larger than $15\,\Mp$, which is known as the Lyth bound \cite{Lyth97}. To obtain this trans-Planckian field value, the Kim-Nilles-Peloso (KNP) 2-flation has been introduced with two axions \cite{KNP05}. It is known recently that the natural inflation is more than $2\sigma$ away from the central value of BICEP2, $(r,n_s)=(0.2,0.96)$. In general, the hilltop inflation gives almost zero $r$.

Therefore, for the \Ude\,hilltop inflation to give a large $r$, one must introduce another field which is called {\em chaoton} because it provides the behavior of $m^2\phi^2$ term at the BICEP2 point \cite{KimHilltop14}. With this hilltop potential, the height is of order $\Mp^4$ and the decay constant is required to be $> 15\,\Mp$. Certainly, the potential energy is smaller than order $\Mp^4$  for
$\phi=[0,f_{\rm de}]$. Since this hilltop potential is obtained from the mother discrete symmetry, such as ${\bf Z}_{10\,R}$, the flat valley at the trans-Planckian $f_{\rm de}$ is possible, for which the necessary condition is given in terms of quantum numbers of  ${\bf Z}_{10\,R}$ \cite{KimHilltop14}.

With the GUT scale vacuum energy during inflation, the reheating temperature after inflation is most likely above $10^{12\,}\gev$. This has led to the conclusion of Ref. \cite{Marsh14} that $f_a$ is around or smaller than $10^{11\,}\gev$. But this study depends on the numerical calculation of \cite{Kawasaki12}. That is the reason for the recent interests on axion string-DW system. A particular interest is the Choi-Kim mechanism \cite{ChoiKimDW85} where several vacua are identified by a subgroup of the Goldstone boson direction. It is particularly useful in string compactification with the anomalous U(1) surviving as a global symmetry down to the axion window scale, because the model-independent axion corresponding to the anomalous U(1) has DW number one \cite{KimDW14}.

\section{Acknowledgments}
JEK is supported in part by the National Research Foundation (NRF) grant funded by the Korean Government (MEST) (No. 2005-0093841) and  by the IBS(IBS CA1310).

\begin{footnotesize}

\end{footnotesize}


\end{document}